# Greatly enhanced intensity-difference squeezing for narrow-band quantum metrology applications

Da Zhang[†], Changbiao Li[†], Zhaoyang Zhang[†], Feng Li[†], Yiqi Zhang[†], Yanpeng Zhang[†,*], Min Xiao[¶,§,*]

[†]Key Laboratory for Physical Electronics and Devices of the Ministry of Education & Shaanxi Key Lab of Information Photonic Technique, Xi'an Jiaotong University, Xi'an 710049, China

[¶]Department of Physics, University of Arkansas, Fayetteville, Arkansas 72701, USA

[§]National Laboratory of Solid State Microstructures and School of Physics, Nanjing University, Nanjing 210093, China

Narrow-band intensity-difference squeezing beams have important applications in quantum metrology and gravitational wave detection. The best way to generate narrow-band intensity-difference squeezing is to employ parametrically-amplified four-wave mixing process in high-gain atomic media. Such IDS can be further enhanced by cascading multiple parametrically-amplified four-wave mixing processes in separate atomic media. The complicated experimental setup, added losses and required high-power pump laser with the increase of number of stages can limit the wide uses of such scheme in practical applications. Here, we show that by modulating the internal energy level(s) with additional laser(s), the degree of original intensity-difference squeezing can be substantially increased. With an initial intensity-difference squeezing of -8.5±0.4 dB using parametrically-amplified-non-degenerate four-wave mixing process in a three-level Λ-type configuration, the degree of intensity-difference squeezing can be enhanced to -11.9±0.4 dB/-13.9±0.4 dB (corrected for losses) when we use one/two laser beam(s) to modulate the involved ground/excited state(s). More importantly, a maximal noise reduction of -9.7±0.4 dB (only corrected for electronic noise) is observed below the standard quantum limit, which is the strongest reported to date in phase insensitive amplification in four-wave mixing. Applying the model to quantum metrology, the signal-to-noise ratio is improved by 23 dB compared to the conventional

Mach-Zehnder interferometer under the same phase-sensing intensity, which is a 14-fold enhancement in rms phase measurement sensitivity beyond the shot noise limit. Our results show a low-loss, robust and efficient way to produce high degree of IDS and facilitate its potential applications.

Traditionally, quantum correlated bright laser beams are generated through parametrically-amplified (PA) optical down-conversion processes in nonlinear optical crystals.[1-4] The produced entangled beams typically have broad spectral width and therefore short coherence time due to the broad phase-matching width in nonlinear crystals.[5-7] Recently, narrow-band bright entangled light beams have been produced through PA four-wave mixing (PA-FWM) process in high-gain atomic media.[8] The IDS between the two beams can reach -8.0 dB without compensating for any system noise or correcting for transmission or detection efficiency.[9] Subsequently, as much as -9.2 dB IDS has been reported by using a pair of high-quantum-efficiency photodiodes and correcting the electronic noise.[10] Several interesting applications of using such narrow-band entangled beams, such as in entangled images,[11,12] FWM slow light,[13] delay of Einstein-Podolsky-Rosen entanglement,[14,15] quantum metrology,[16-21] have all been experimentally demonstrated. In order to further increase the degree of IDS, the technique of cascading more stages of PA-FWM process has been employed. In one experiment, a second PA-FWM process in a separate atomic vapor cell was used to enhance the IDS from $(-5.5\pm0.1)$ dB/$(-4.5\pm0.1)$ dB to $(-7.0\pm0.1)$ dB.[22] Similarly, enhanced continuous-variable squeezed states have also been realized by cascading two PA-down-conversion processes using two separate nonlinear crystals.[23] Ultimate enhancement limit reachable by using more stages of such cascade setups can be theoretically derived.[24] Although the cascading technique is conceptually simple to consider, the added complications in the experimental setup and the required high-power pump laser with the increase of number of stages, as well as maintaining the phase coherence, between different stages will severely limit the broad applications in using such generated high degree of IDS.

Here we implement a totally different approach to enhance the IDS produced from the two correlated light beams generated from the PA-FWM process in a three-level Λ-type atomic system. Since the degree of IDS is mainly determined by the high optical gain realizable in the PA-FWM process[8], we set to find an efficient way to enhance the optical gain in the same atomic system. In our experiments, the dressing fields, to significantly improve the conversion efficiency in four-wave mixing as a major benefit of constructive interference between different transition probability amplitudes,[25] which produces efficient higher-order multi-wave mixing processes.[26-28] This scheme of enhancing parametric gain, and therefore the generated IDS, in the system by modulating the internal states of a multi-level atomic system has certain obvious advantages over using separate cascading stages in enhancing IDS. The first is the lower optical path loss. The second is higher squeezing limit with less vacuum losses because of one stage rubidium cell. The third is that the dressing field can improve the noise figure of system and make it close to the quantum limit. At the same time, the degree of IDS cannot only be greatly enhanced, but also suppressed by simply varying the frequency detunings of the additional driving fields. Compared with the case of a single cell, our model has lower pump power limit and higher gain saturation limit because of degenerate multi-wave mixing process. These merits will greatly facilitate the potential applications of such IDS light sources in entanglement imaging,[11,12] quantum metrology,[16-21] quantum communication,[29-32] and quantum information processing.[29,32]

## RESULTS AND DISCUSSIONS

**Theoretical model**

Let us first consider the three-level Λ-type sub-system (involves levels $|0\rangle \leftrightarrow |2\rangle \leftrightarrow |1\rangle$), as shown in Figure. 1(b). A strong pump beam $E_1$ (with frequency $\omega_1$, $\mathbf{k}_1$, Rabi frequency $G_1$, vertical polarization) is tuned to couple the D2 line (780 nm) transition and a weak beam $E_2$ ($\omega_2$, $\mathbf{k}_2$, Rabi frequency $G_2$,

horizontal polarization) works as a probe field. The detuning $\Delta_i=\Omega_i-\omega_i$ is defined as the difference between the resonant transition frequency $\Omega_i$ and the laser frequency $\omega_i$ of $E_i$. With the frequency of $E_1$ tuned far away from the resonances, this system forms the standard PA-non-degenerate-FWM configuration to satisfy the phase-matching condition $\mathbf{k}_S^F + \mathbf{k}_{aS}^F = 2\mathbf{k}_1$ (as shown in Figure. 1c1, and produces narrow-band IDS between the parametrically amplified probe (anti-Stokes) and conjugate (Stokes) beams. The generated IDS mainly depends on the gain factor $T_F$ in the anti-Stokes channel. This nonlinear gain factor $T_F$ can be modified by multiple parameters in multi-level coherent atomic systems.

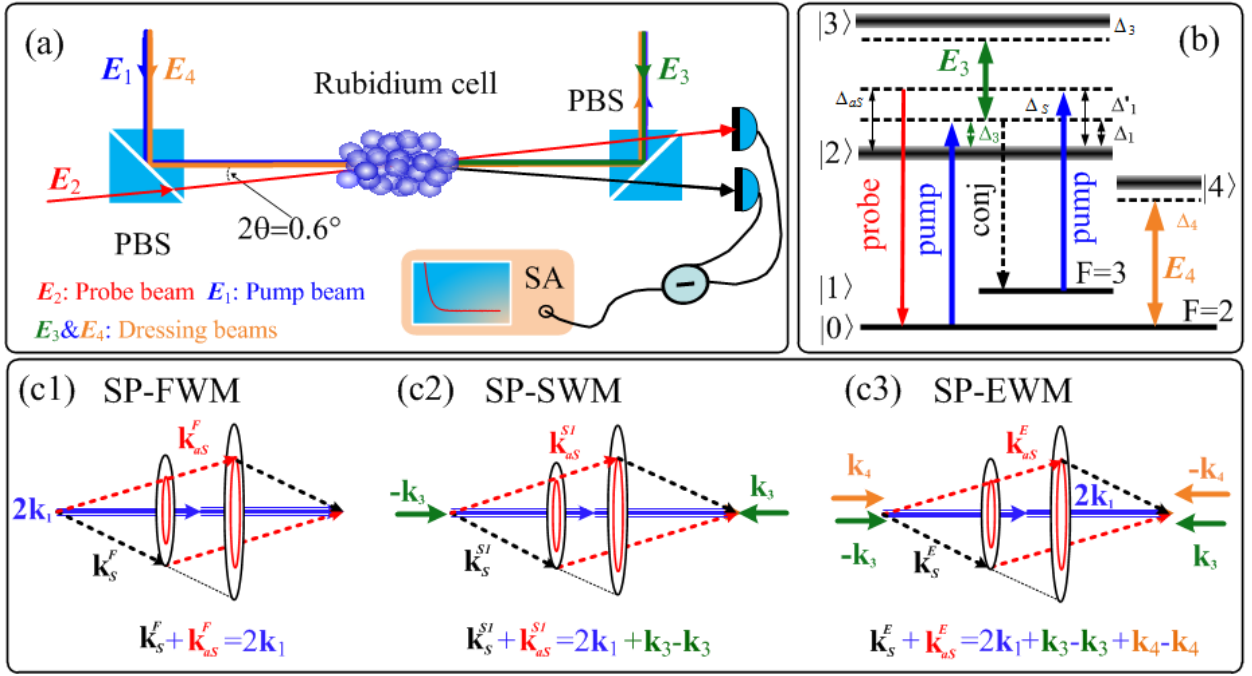

**Figure. 1** (a) Experimental setup. PBS: polarizing beam splitter; SA: spectrum analyzer. (b) Energy level diagram of the (Λ-type (|0⟩↔|1⟩↔|2⟩)) rubidium atomic system with an $E_3$ ladder-type dressing (between levels |2⟩ and |3⟩) and an $E_4$ V-type dressing (between levels |0⟩ and |4⟩) simultaneously. (c1)-(c3) Phase-matching conditions for the spontaneous parametric FWM (SP-FWM), SP-SWM1 and SP-EWM processes, respectively.

Next, let us turn our attention to energy-level modulation (one-beam dressing) with an additional laser beam $E_3$ ($\omega_3$, $\mathbf{k}_3$, $G_3$, $\Delta_3$) in the ladder-type dressing scheme (Figure. 1b). The interaction Hamiltonian of this $E_3$-dressed PA-FWM process can be expressed as (with all pump and dressing fields treated as classical fields):

$$H = i\hbar\kappa_{F1}^{1d}\hat{a}^+\hat{b}^+ + H.c, \tag{1}$$

where $\hat{a}^+$ and $\hat{b}^+$ are the boson creation operators acting on the electromagnetic excitation of the anti-Stokes and Stokes channel, $\kappa_{F1}^{1d} = |-i\varpi_{s/as}^2 \chi_{F1}^{1d(3)} E_1^2 / 2k_{s/as}|$ is the pumping parameter for the $E_3$-dressed PA-FWM process, which depends on the nonlinear susceptibility tensor $\chi_{F1}^{1d(3)}$ and the pump-field amplitude. $\varpi_{s/as}$ is the central frequency of generated Stokes or anti-Stokes signal. In the dressed-state picture, the third-order nonlinear susceptibility tensor can be expressed as $\chi_{F1}^{1d(3)} = |N\mu_{20}\mu_{21}\rho_{F1(s/as)}^{1d(3)} / \varepsilon_0 \hbar E_1^2 G_{(s/as)}|$, where $\rho_{F1(s/as)}^{1d(3)}$ is the corresponding $E_3$-dressed density-matrix element which can be obtained from eq S2 and S3 in Supplementary Note 1. Here, we define the $E_3$-dressed nonlinear gain coefficient (related to the matrix element $\rho_{F1(s/as)}^{1d(3)}$) as $T_{F1}^{1d} = \cosh^2(\kappa_{F1}^{1d} L)$. The modified degree of IDS in this $E_3$-dressed PA-FWM is given by:

$$Sq_{F1}^{1d} = -Log_{10}(2T_{F1}^{1d} - 1). \tag{2}$$

This single-beam ($E_3$) modulated FWM process can be decomposed into co-existing FWM and six-wave mixing1 (SWM1, with phase-matching condition $\mathbf{k}_S^{S1} + \mathbf{k}_{aS}^{S1} = 2\mathbf{k}_1 + \mathbf{k}_3 - \mathbf{k}_3$, Figure. 1c2) processes[27,28]. Therefore, $\kappa_{F1}^{1d}$, $\chi_{F1}^{1d(3)}$, and $T_{F1}^{1d}$ are all greatly modified by the dressing field $E_3$ in eq S2 and S3 of Supplementary Note 1. With increased nonlinear gain factor $T_{F1}^{1d}$, $Sq_{F1}^{1d}$ can be greatly enhanced. Similarly, we can obtain the one-beam $E_4$ ($\omega_4$, $\mathbf{k}_4$, $G_4$, $\Delta_4$) dressed PA-FWM gain factor $T_{F2}^{1d}$ in the similar way and, therefore, achieve an enhanced IDS ($Sq_{F2}^{1d}$) in such V-type dressing configuration (Figure. 1b), which corresponds to co-existing the FWM and another SWM2 (with $\mathbf{k}_S^{S2} + \mathbf{k}_{aS}^{S2} = 2\mathbf{k}_1 + \mathbf{k}_4 - \mathbf{k}_4$) processes.

Finally, when the $E_3$ and $E_4$ beams turn on simultaneously (Figure. 1b), a two-beam dressed PA-FWM configuration is formed, which generates co-existing FWM, SWM1, SWM2 and eight-wave mixing (EWM, with phase-matching condition $\mathbf{k}_S^E + \mathbf{k}_{aS}^E = 2\mathbf{k}_1 + \mathbf{k}_3 - \mathbf{k}_3 + \mathbf{k}_4 - \mathbf{k}_4$) processes (Figure. 1c3). We can obtain the two-beam-modulated gain factor $T_F^{2d}$ and IDS $Sq_F^{2d}$ in this five-level atomic system. The expression for the modulated third-order nonlinearity (See details in Supplementary Note 1) clearly reveals why the gains of the modified PA-FWM systems can be greater than the original PA-FWM process, and therefore the enhanced IDS.

However, there exist inevitable losses in the system, such losses with vacuum field coupling will make a limit value of the squeezing.[24] Let's assume the dressing field has extended to N. Considering the loss of the system, the vacuum coupling term (loss term) $m\hat{c}_1$ and $m\hat{c}_2^+$ will occur at Heisenberg evolution equation of operator (eq S4 and S5 in Supplementary Note 1). The IDS of n-dressing PA-FWM can be modified as $Sq_F^{nd} = Log_{10}[(1+m^2)/(2T_F^{nd}-1)]$. With the increase of separate cascading stages, loss term is liable to rapid accumulation,[22,24] lead to the limit of squeezing. In the cascade system, the squeezing limit depends on the added losses and gain saturation of the system.[22] However it also mainly depend on the gain saturation in our system with one stage rubidium cell, which means we have a higher limit value. The theory analyses and experiment results show our method has a very good scalability.

Note that we have used different orders of multi-wave mixing (MWM) processes to describe the dressed PA-FWM processes, which give a clear physical picture for the complicated situations and is valid under certain approximations on the dressing fields.[33,34] With the clear decompositions of the dressed-state formulism for the multi-beam-dressed PA-FWM, we can better identify the contributions of the modified IDS from different wave-mixing processes. Such methods show a robust and efficient way to produce high degree of IDS.

**Enhanced IDS with one dressing field (either $E_3$ or $E_4$)**

There are two ways to dress the Λ-type three-level ($|0\rangle \leftrightarrow |1\rangle \leftrightarrow |2\rangle$) system, one by applying $E_3$ between levels $|2\rangle$ and $|1\rangle$ (i.e., ladder-dressing configuration) and another by applying $E_4$ between levels $|0\rangle$ and $|4\rangle$ (V-dressing configuration), as shown in Figure. 1b, which modify the original PA-FWM process differently, and therefore provide different enhancement factors for IDS. In the following, we consider their effects separately.

Figure 2 shows PA-FWM signals in the probe and corresponding conjugate channels in the three-level Λ-type $^{85}$Rb atomic system, respectively, with and without the $E_3$ beam. With $E_3$ off and the pump field detuning $\Delta_1$ detuned to ~1.12 GHz, we scan the probe field $E_2$ over 8 GHz across D2 line and observe a number of features in transmission and its conjugate channels, as shown in Figure. 2a1 and 2a3, respectively. When $E_3$ turns on, as shown in Figure. 1c2 the one-beam dressing FWM (coexisting FWM+SWM)[27] signal in Figure. 2a2 gets stronger than that in Figure. 2a1, which indicates an enhanced FWM process. The $E_3$-dressed PA-FWM signal is enhanced due to the constructive interference between FWM and SWM1, satisfying the dressed enhancement condition $\Delta_1+\Delta_3 \pm \sqrt{\Delta_3^2 + 4|G_3|^2}/2 = 0$. Similar to the probe channel, the corresponding conjugate signal is also enhanced due to the existing PA-SWM1 process, as shown in Figure. 2a4. So, the dressed gain coefficient $T_{F1}^{1d}$ is enhanced compared to the gain coefficient $T_F$ without the $E_3$ beam (Figure. 2a3).

The measured IDS of the PA-FWM signal (Figure. 2b2) is -3.6±0.4 dB below the normalized SQL. With the dressing field $E_3$ on, the measured IDS of the $E_3$-dressed FWM signal (Figure. 2b3) is about -6.1±0.4 dB below the SQL, which indicates that the degree of IDS ($Sq_{F1}^{1d}$ in eq 2) is significantly increased by the enhanced nonlinear gain coefficient $T_{F1}^{1d}$ due to the $E_3$ dressing effect, as shown in Figure. 2b3. Furthermore, one can infer IDS of the pure PA-SWM1 to be -2.8±0.4 dB.

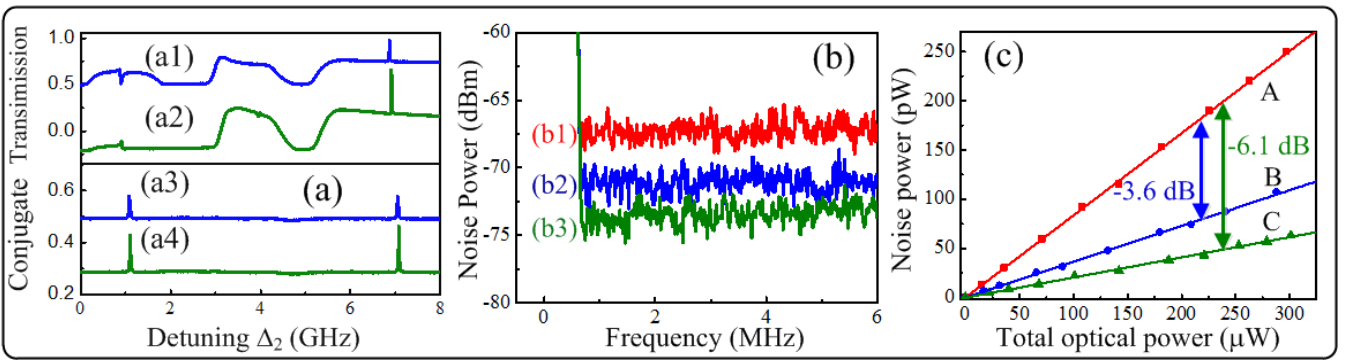

**Figure.** 2 Measured noise spectra and modulated IDS with $E_3$-beam dressing. (a) Measured probe transmission ($E_{aS}$) and corresponding conjugate ($E_S$) signal versus probe frequency detuning. (a1) and (a3) are with $E_3$ off while (a2) and (a4) are with $E_3$ on, where as $\Delta_1$=1.12GHz, $\Delta_3$=-1GHz. (b) Relative intensity noise levels versus spectrum analyzer frequency. (b1)-(b3) SQL, FWM, and dressed FWM (FWM+SWM1) when $E_3$ is applied, respectively. (c) Relative intensity noise power at different total optical power for (A) SNL (diamonds); (B) FWM (circles); (C) single-dressing FWM (triangles). All these three noise power curves fit to straight lines. The electronic noise floor and background noise are subtracted from all of the traces and data points.

To better show the squeezing enhancement as predicted by the theory, we measure the relative intensity noise power for the FWM (curve B in Figure. 2c) and single-dressing FWM (curve C) at 1 MHz as a function of the total optical power impinging on the photodetectors. We also record the noise powers of a coherent beam at different optical powers using the SNL measurement method described above (curve A). We can see that the ratios of the slopes for curve B/A and curve C/A are 0.437±0.038 and 0.245±0.038, respectively, which indicate the degrees of squeezing of the FWM and single-dressed FWM to be about -3.6±0.4 dB and -6.1±0.4 dB, respectively. The optical path transmission is 80%, resulting in a total detection efficiency of 64.8%; the uncertainty is estimated at 1 standard deviation. The inferred degrees of squeezing for the FWM and $E_3$-dressed FWM beams are -8.5 dB and -11.0 dB after corrected for losses, respectively. To unify the standards, here we report the corrected values without special explanation.

Next, we consider the case with $E_4$-dressed PA-FWM instead of $E_3$. Figure. 3a2 and 3c2 show that, compared to the original PA-FWM (Figure. 3a1 and 3c1), the $E_4$-dressed FWM signal in the probe channel can be either enhanced or suppressed. The field $E_4$ dresses the ground state $|0\rangle$ and creates the dressed states $|G_{4\pm}\rangle$. Thus, due to the fulfillments of dressed enhancement and suppression conditions, i.e., $\Delta_1 - \Delta_4 \pm \sqrt{\Delta_4^2 + 4|G_4|^2}/2 = 0$ and $\Delta_1+\Delta_4$=0, the PA-FWM signal can be either enhanced or suppressed[26,27]. The increase or decrease of the probe and conjugate field intensities (Figure. 3a2 and c2) is caused by constructive or destructive interference between the generated FWM and SWM2 fields. Therefore, the corresponding dressed gain $G_{F2}^{1d}$ becomes large or small accordingly.

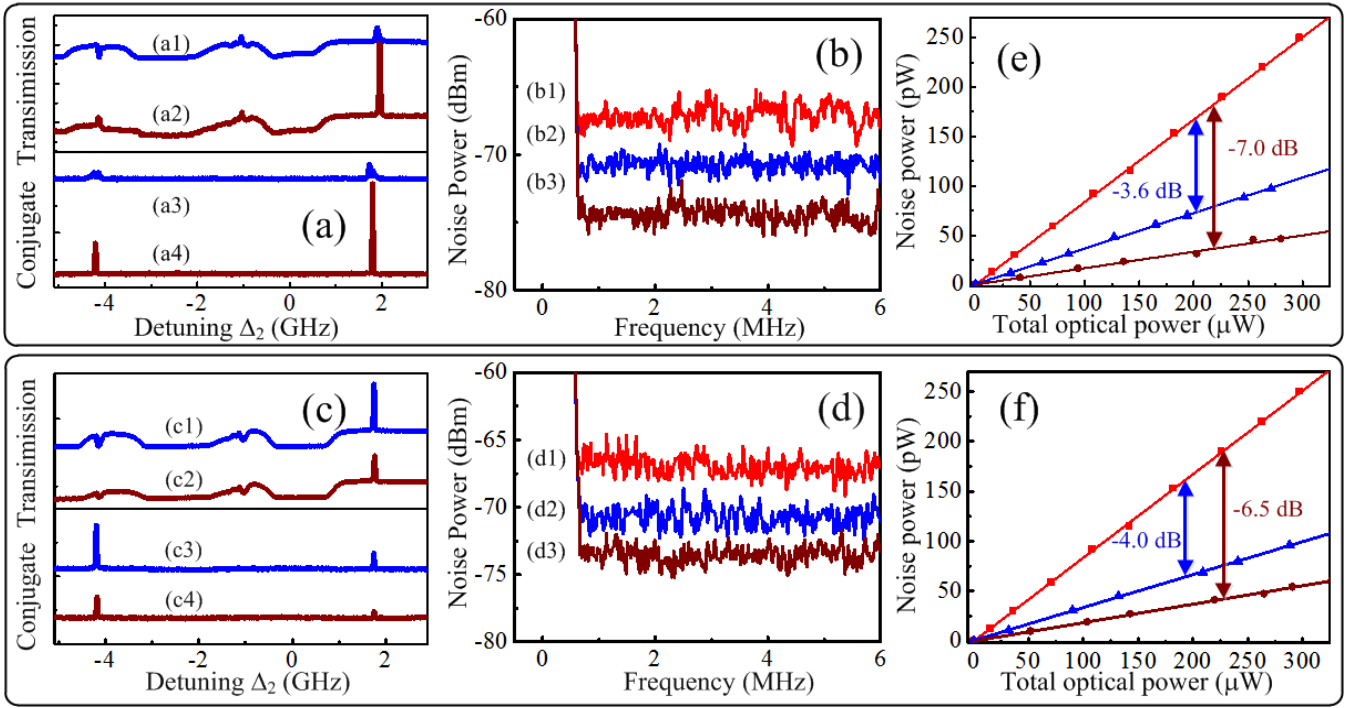

**Figure.** 3 Measured noise spectra and modulated IDS with $E_4$-beam dressing. (a1) and (a3) are with $E_4$ off. (a2) and (a4) with $E_4$ on, and $\Delta_1$=1.12GHz, $\Delta_4$=-1GHz. (b1)-(b3) SQL, FWM, and dressed FWM (FWM+SWM2) when field $E_4$ is applied, respectively. (c) and (d) $\Delta_1$=1.15 GHz, $\Delta_4$=1.15 GHz, respectively. (e) Relative intensity noise power at different total optical power for SNL (diamonds); FWM (triangles); enhanced $E_3$-dressing FWM (circles). (f) It's like (e) except that it is suppressed $E_3$-dressing FWM. All these noise power curves fit to straight lines. The electronic noise floor and background noise are subtracted from all of the traces and data points.

Figure. 3(b) and 3(d) depict the measured IDS of $E_4$-dressed FWM, corresponding to enhanced (Figure. 3b) and suppressed (Figure. 3d) conditions, respectively. The measured degree of IDS ($Sq_{F'}^{1d}$) for this $E_4$-dressed FWM (Figure. 3b3) is -11.9±0.4dB, which is much larger than that of the original PA-FWM (-8.5±0.4dB) (Figure. 3b2). The relative intensity noise power for FWM (triangles) and $E_4$-dressed FWM (circular) are shown in Figure. 3e with the change in total optical power. They are similar to Figure. 2c, we will not repeat here. Moreover, the inferred -3.7dB IDS for the PA-SWM2 process is larger than the -2.8 dB IDS for the PA-SWM1 process with $E_3$ dressing shown in Figure. 2b3. The reason is that $E_4$ has effects of both population transfer and dressing (eq S23 in Supplementary

Note 3) while $E_3$ only has the dressing gain ($T_{F1}^{1d}$), resulting in a much larger total nonlinear gain $T_{F'}^{1d}$ for the $E_4$-dressed case.

However, due to the existed population gain $T^p$ for the V-type dressing scheme, the $E_4$-dressed PA-FWM cannot realize a suppression of IDS below the original PA-FWM value. In fact, under the suppressed gain condition (destructive interference between the FWM and SWM2 fields), the measured IDS of the $E_4$-dressed PA-FWM is still as large as -13.5±0.4 dB (Figure. 3d3). Figure. 3f shows the corresponding dependencies of SNL, FWM and $E_4$-dressed FWM signals on optical power.

**Enhanced IDS with two dressing fields**

Finally, let's consider the case with both $E_3$ and $E_4$ dressing fields on at the same time for the five-level system, as shown in Figure. 1b, with phase-matching conditions given in Figure. 1c3. Figure. 4a and 4b show modified PA-FWM signals in the probe and conjugate channels, respectively, when different beam(s) are blocked. As the frequency detunings of $E_3$ and $E_4$ set to be -0.9GHz and 0.95GHz, the intensity of dressed PA-FWM is expected to increase relative to Figure. 4a1, as shown in Figure. 4a2 and 4a3, respectively. Similar to Figure. 2a2 and 3a2, two newly generated PA-SWM signals, i.e., SWM1 and SWM2, increase in Figure. 4a2 and a3. Particularly, with $E_3$ and $E_4$ both on simultaneously, a two-beam-dressed FWM signal in Figure. 4a4 is greatly enhanced, which is the mixture of one pure PA-FWM, two SWMs and one EWM processes. At the same time, the intensity of the two-beam-dressed conjugate signal ($E_S$) is also changed accordingly, as shown in Figure. 4b. So, the two-beam dressing gain $T_F^{2d}$ is significantly enhanced.

Figure. 4(c) presents the measured degrees of IDS for modified PA-FWM signals in Figure. 4a and 4b. First, with all external dressing fields ($E_3$ and $E_4$) blocked, and the IDS of pure PA-FWM is measured to be -8.5±0.4 dB (Figure. 4c2). The curve (c1) gives the SQL. When either $E_3$ or $E_4$ is on, the measured IDS of one-beam-dressed FWM is -9.9±0.4 dB (Figure. 4c3) or -11±0.4 dB (Figure. 4c4), respectively. When both dressing fields ($E_3$ and $E_4$) are on at the same time, the measured IDS of two-beam-dressed FWM (Figure. 4c5) reaches -13±0.4 dB. Their corresponding dependencies of SNL; FWM; $E_3$-dressed FWM; $E_4$-dressed FWM, and double-dressed FWM signals on optical power are shown in Figure. 4d. After fitting all these five noise power curves to straight lines, we find that the ratios of slopes between curve B,C,D,E and A equal to 0.436±0.038, 0.316±0.038, 0.245±0.038, and 0.126±0.038, respectively, which shows that the degree of IDS of the twin beams are about -3.6±0.4 dB, -5.0±0.4 dB, -6.1±0.4 dB and -9.0±0.4 dB, respectively. This largely increased degree of IDS is caused by the enhancement in two-beam-dressed PA-FWM gain with coexisting and constructive interfered PA-FWM, SWM and EWM processes in the system. The inferred degree of IDS for the pure PA-EWM is -2.3±0.4 dB ($Sq_{F'}^{2d}$ in eq S10 of Supplementary Note 1). Furthermore, the IDS of double-dressed FWM can reach -9.7 dB after corrected for electronic noise which is slightly larger than the best results in FWM[10]. The inferred squeezing value at the end of the atomic medium, corrected for detection efficiency, is better than -13.9 dB with weak probe injection. In our method, the total degree of IDS can be easily controlled and modulated by adjusting the frequency detunings of the dressing fields.

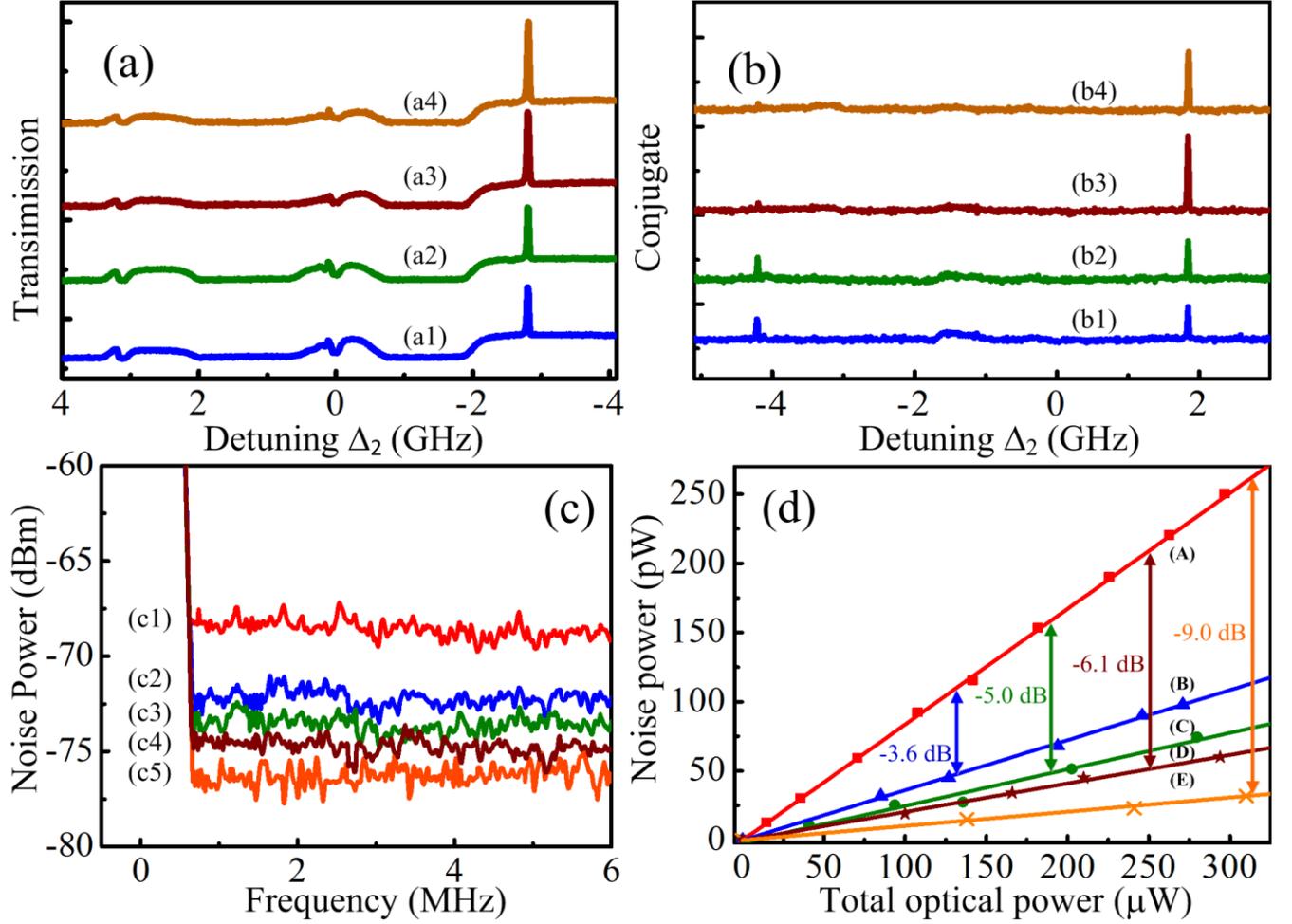

**Figure.** 4 Measured noise spectra and enhanced IDS with both $E_3$ and $E_4$ beams on simultaneously. (a) Measured probe transmission signal ($E_{aS}$) versus the probe detuning. (a1) $E_3$ and $E_4$ off; (a2) $E_3$ on; (a3) $E_4$ on; (a4) $E_3$ and $E_4$ both on, and set $\Delta_1$=1.12GHz, $\Delta_3$=-0.9 GHz and $\Delta_4$=0.95 GHz. (b1)-(b4) corresponding conjugate signal ($E_S$) of (a1)-(a4), respectively. (c) Relative intensity noise spectra versus spectrum analyzer frequency. (c1)-(c5) SQL, FWM, $E_3$-dressed FWM, $E_4$-dressed FWM, and $E_3$- &$E_4$-dressed FWM (FWM+SWM+EWM), respectively. (d) Relative intensity noise power at different total optical power for (A) SNL (diamonds); (B) FWM (triangles); (C) $E_3$-dressing FWM(circles), (D) $E_4$-dressing FWM (pentagon), and (E) double-dressing FWM (cross). All these five noise power curves fit to straight lines. The electronic noise floor and background noise are subtracted from all of the traces and data points.

Replacing the FWM beam splitter with double-dressed FWM model,[20] we can obtain a non-conventional interferometer with higher sensitivity. As is proved by theoretical analyses,[20,36] the interference fringe of non-conventional interferometer is enhanced by $2T_F^2$ compared with a conventional interferometer for the same phase-sensing intensity conditions. In our case, the improvement of the signal-to-noise ratio is 23 dB with the initial injection of 15 μW. Furthermore, we have the phase measurement sensitivity as 14 in root-mean-square value.

## CONCLUSIONS

In conclusion, we have observed the greatly enhanced IDS for single- and double-beam modulated PA-FWM processes in the same hot atomic system. Compared to the simple PA-FWM case (with -8.5 dB IDS), the degrees of IDS for $E_3$- are $E_4$-modulated PA-FWM processes are measured to be -9.9 dB and -11.9 dB, respectively. The degree of IDS for the two-beam-dressed PA-FWM process gets up to -13.9 dB, which indicates that the generated higher-order PA-MWM processes contribute to the total parametric gain and therefore the quantum noise suppression (or enhanced IDS). Under different dressing frequency detunings, the generated high-order nonlinear signals can interfere either constructively, which enhances the total parametric gain, or destructively, which reduces the total gain. Our current experiment demonstrates a robust and efficient way to produce high degree of IDS on an integrated platform which can find potential applications in quantum metrology and gravitational wave detection[21].

## METHODS

The five relevant energy levels are $5S_{1/2}$, $F=2$ ($|0\rangle$), $5S_{1/2}$, $F=3$ ($|1\rangle$), $5P_{3/2}$ ($|2\rangle$), $5D_{5/2}$ ($|3\rangle$), $5P_{1/2}$ ($|4\rangle$) in $^{85}$Rb, as shown in Figure. 1b. Levels $|0\rangle\leftrightarrow|2\rangle\leftrightarrow|1\rangle$ form the basic Λ-type three-level system. We use light of 500 mW from a CW Ti:Sapphire laser as the 780 nm pump beam ($E_1$), and another light up to 0.2 mW from an external cavity diode laser as the 780 nm probe beam ($E_2$). They couple with the Λ-type atomic system in a naturally abundant rubidium vapor cell by a polarizing beam splitter (PBS). The vapor cell is wrapped with μ-metal sheets to shield stray magnetic field from environment and heated to 125°C to provide an atomic density of $3\times10^{13}$cm$^{-3}$. $E_2$ propagates in the same direction as $E_1$

with a small angle of 0.26°. These two laser beams form the standard double-Λ configuration and produce the PA-FWM IDS. When $E_4$ (795 nm, 4 mW) is added onto $E_1$ (in the same direction) and $E_3$ (776 nm, 8 mW) counter-propagates with $E_1$, they establish two electromagnetically induced transparency windows in the system and significantly dressing the original PA-FWM process.[28,35] The dressing fields of 776 and 795nm are provided by two Toptica lasers. Their frequencies are locked but phases are unlocked. The spatial alignments of the beams are shown in Figure. 1a. The output probe and conjugate beams are detected by two balanced photodetectors. The difference of the two detected signals is sent to a radio frequency spectrum analyzer with a resolution bandwidth of 300kHz and a video bandwidth of 10 kHz. All intensity difference measurements presented in this paper are taken at an analysis frequency of 1 MHz.

Subsequently, the noise spectra of the relative intensities between the probe and conjugate channels are measured. First, $E_1$ passes through an acousto-optic modulator (AOM) of 1.5 GHz twice to have a frequency difference of 3 GHz. The beam is then injected into the probe channel and the field $E_2$ is off. To calibrate the standard quantum limit (SQL) for the total optical power arriving at the photodetectors, a coherent beam with the same power is split by a 50/50 beam splitter, and directing the resulting beams into a balanced and amplified photodetector with a transimpedance gain of $10^5$ V/A and 81% quantum efficiency.

## AUTHOR INFORMATION

Corresponding Authors *E-mail: ypzhang@mail.xjtu.edu.cn. *E-mail: mxiao@uark.edu.

**Notes** The authors declare no competing financial interest

## ACKNOWLEDGMENTS

This work was supported by the National Key R&D Program of China (2017YFA0303700), the National Nature Science Foundation of China (61308015, 11474228), and the Key Science and